# Thematic analysis of student perceptions of resources and demands experienced in introductory physics

Avital Pelakh[1,2], Melanie L. Good[3], Eric Kuo[4], Michael Tumminia[2,5], Nabila Jamal-Orozco[1,2], Amy Adelman[1,2], Jordann Antoan[1,2], Brian Galla[2,5], and Timothy J. Nokes-Malach[1,2]

[1]*Department of Psychology, University of Pittsburgh, Pittsburgh, PA 15260, USA*
[2]*Learning Research and Development Center, University of Pittsburgh, Pittsburgh, PA 15260, USA*
[3]*Department of Physics and Astronomy, University of Pittsburgh, Pittsburgh, PA 15260, USA*
[4]*Departments of Physics and Curriculum & Instruction, University of Illinois at Urbana-Champaign, Urbana, IL 61820, USA*
[5]*Department of Health and Human Development, University of Pittsburgh, Pittsburgh, PA 15260, USA*

The current work aims to better understand student course experiences for those who reported negative perceptions in introductory physics. We conducted semi-structured interviews with 24 students who reported negative perceptions of their class on a screening survey. Participants were asked to share general reflections on challenges and successes they experienced, as well as their reflections on specific aspects of the course (e.g., experiences with instructors and peers). Interview transcripts were then coded to identify the types of experiences students reported, whether they were experienced as positive or negative, as well as the themes and features associated with those experiences. Experiences with the classroom, course structure, instructors, and exams were most frequently reported as negative. Experiences with peers, help-seeking, course curriculum, and specific learning activities were the most positive, though only experiences with peers had more positive reports than negative. We then used a *resources vs. demands* framework [Soc Personal Psychol Compass 7, 637 (2013)] to interpret the common instructional, cognitive, and motivational themes and features reported across multiple contexts. We discuss the implications of the results for theory and practice.

## I. INTRODUCTION

There has been a growing research focus on motivation and engagement in undergraduate introductory physics courses. Researchers have used self-report surveys to show that many physics students report low self-efficacy [1–3], physics identity [4,5], and belonging [6], which is associated with negative emotions [7], disengagement [2], and attrition [6]. Moreover, several studies have shown that self-efficacy, interest, and sense of belonging shift negatively after physics instruction in introductory course sequences, suggesting a negative impact of course experiences on these motivational beliefs [6,8,9]. Further, these outcomes may disproportionally impact students with historically excluded identities in physics (e.g., students of color, women) who often face additional barriers such as stereotypes [10,11] and bias [12,13]. To understand the experiences of students who view physics negatively, we conducted semi-structured interviews with participants who self-reported below-average motivation and negative feelings and attitudes toward their physics course. We developed a coding rubric to describe the types of experiences and how common they were, whether they were negative or positive, and identified common themes and features across them. We also report which aspects of the experiences tend to co-occur together to better describe the multifaceted nature of student experiences in the course. We then apply a theoretical framework of resources and demands [14] to interpret the themes and features in terms of classroom ecologies involving instruction, cognition, and motivation.

The large-enrollment introductory courses that have been linked to negative student experience [15,16] are also those most commonly encountered by students due to requirements for physics and non-physics majors alike, making them a high-priority research context. However, much of the work that has been done in these courses has focused primarily on quantitative measures of either cognitive [17–19] or motivational [1,2,12,20,21] engagement, not both, and therefore risks overlooking important nuances and connections. Studies leveraging qualitative methods have been better able to weave disparate aspects of student experience into a holistic understanding [22,23], but rarely focus on large-enrollment courses. Qualitative study of large-enrollment introductory courses can reveal the nature of relationships between cognition, motivation, and instruction, and can help identify aspects of these particular courses that can be especially difficult for students.

A primary contribution of the current work is a focus on students with more negative beliefs, feelings, and attitudes toward their physics course (relative to their peers). Students were given an opportunity to self-describe both their positive and negative experiences in introductory physics, helping identify instructional and other course features that affect (positively and negatively) students who need the most support [24]. Further, focusing on student beliefs, feelings, and attitudes about physics rather than particular identities (e.g., women, students of color) avoids conflating demographic variables with psychological ones. While studies focused on social identities have revealed important insights about systemic inequities and associated effects [1,2,4,7,13,25], work that centers psychological experience can help identify the mechanisms associated with observed social group disparities.

As a second contribution, we apply a conceptual framework for interpreting those student experiences. The conceptual framework is rooted in the *biopsychosocial model of challenge and threat* [14], which describes experiences in motivated performance contexts in terms of an interaction between situational *demands* (e.g., task difficulty) and the *resources* an individual has to meet those demands (e.g., prior knowledge, motivation, social support). Using information about aspects of the course that students experienced both positively and negatively, we identify the demands and resources that students perceived. We discuss times when instructional intent clashed with student perceptions and hidden demands that may be underappreciated or inadequately supported by instructors and course structures. We also discuss aspects of instruction and the environment that may support the development of student resources in the class.

### A. Classifying Learning Situations in Terms of Resources and Demands

We draw on the biopsychosocial model of challenge and threat [14,26,27], because it can help explain behaviors in stressful or competitive environments [28], has previously been applied to understand learning and academic performance [29,30], and provides a simple classification structure that can be applied to nearly any situation encountered by students. The biopsychosocial model is often used to study unique patterns of physiological, behavioral, and affective responses [26,31–36]. However, we will not focus on these measures in the current study; instead, we will use the model as a conceptual framework for thinking about student experiences and how they arise and will henceforth refer to it as the *resources vs. demands framework*.

The resources vs. demands framework is built around the idea that our experiences and actions in the context of motivated performance situations (e.g., challenging homework assignment) are informed by our cognitive appraisals of the *demands* of the situation (e.g., perceptions of difficulty, time and effort required), as well as the *resources*[1] (e.g., prior knowledge and skills, motivation) we have available to meet those demands. A student can be engaged and motivated to perform, because they have self-relevant goals (e.g., to gain expertise, demonstrate competence). If the student's resources are sufficient to meet or exceed the situational demands, they experience a state of *challenge*, which is associated with positive emotions and a tendency to actively engage in the situation. However, if perceived demands exceed resources, they experience the situation as a *threat*, which is associated with negative emotions and a tendency toward avoidance. Challenge and threat are polar ends of a continuum on which experiences can fall at any point. The model is also iterative: challenge and threat appraisals in the present can be influenced by past appraisals and can influence future appraisals. A conceptual model of the resources vs. demands framework is depicted in Fig. 1.

[1] Note that our use of *resources* is rooted in the biopsychosocial model and is distinct from the cognitive and epistemological resources [81] commonly referenced in PER.

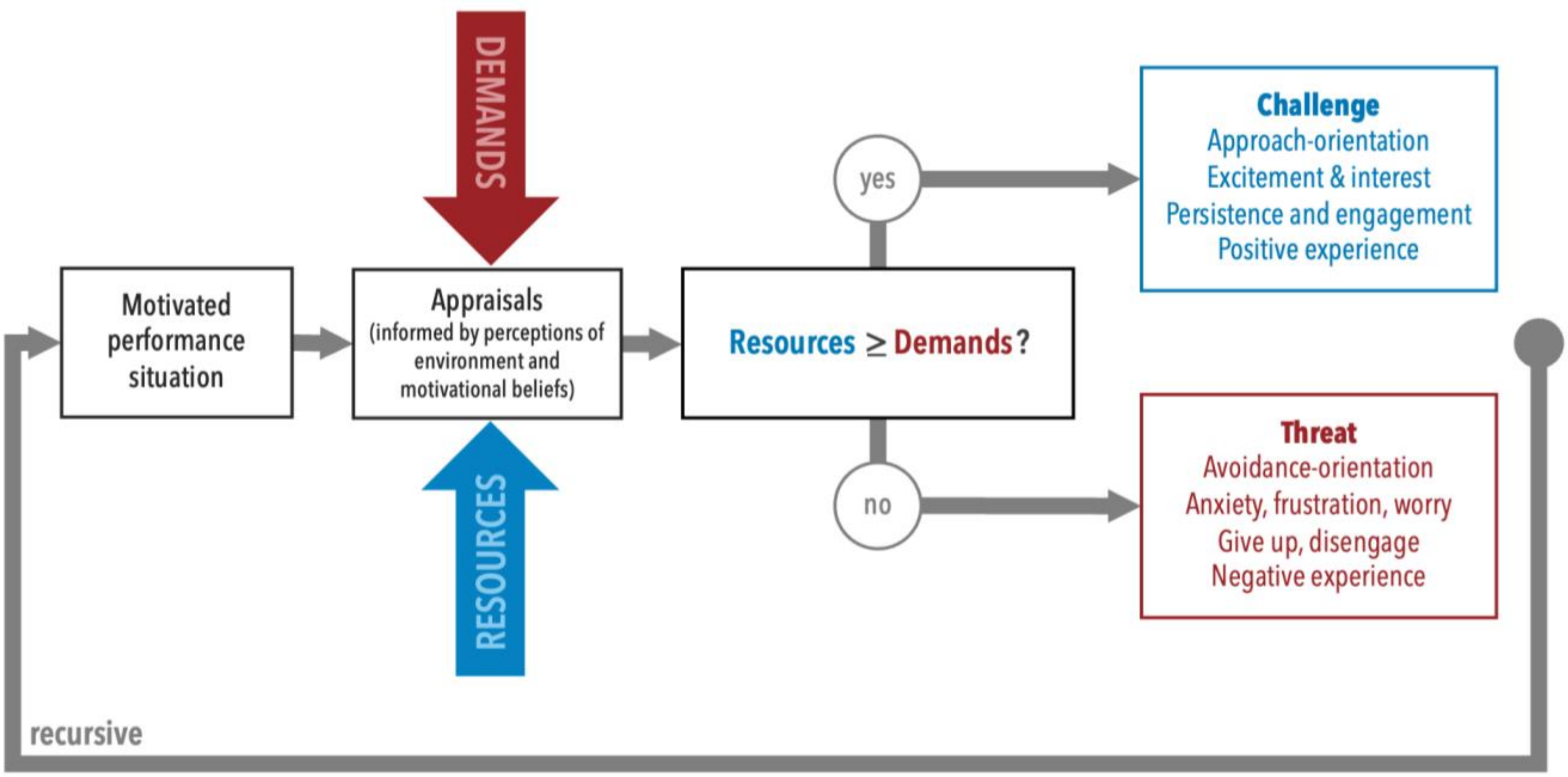

FIG. 1. Conceptual model of the resources vs. demands framework.

This model is useful for understanding course-related experiences, because it can be flexibly adapted to fit a broad array of situations students experience across a wide range of time intervals, from a single exchange with a peer or instructor to their experience with a course as a whole. We use the term "situation" to describe the set of circumstances and conditions existing at a particular time interval and/or place, and "experience" to describe the subjective perception and interpretation of those circumstances, as someone actively lives through them. FIG. 1 models students' experiences of perceived challenge and threat as arising from an appraisal of a situation.

Our theoretical assumption that students' appraisals of resources and demands are created, at least in part, through their interactions with their learning environment. Therefore, by adding, removing, or modifying aspects of the environment, we can shift the balance of resources and demands[2] and create the conditions under which students are more likely to have positive learning experiences [37]. Taking inventory of demands and resources allows us to critically evaluate their connection to physics teaching and learning. For example, demands can be assessed based on their relation to course learning objectives and practical necessity. Resources can be evaluated based on how well they align with specific demands. Then, instructional features can be evaluated based on whether they create demands that are essential for learning, and whether they can either help increase resources (e.g., intervention to promote particular motivational beliefs or additional practice materials).

### B. Current Work

We recruited college students who reported negative feelings and perceptions related to their introductory physics course to participate in semi-structured interviews. In the interviews, we asked participants to describe which aspects of their course they found most challenging, which were helpful, and how those experiences related to their experiences, goals, and beliefs. Student interviews were transcribed and coded to identify the types of experiences students reported and whether the affective valence of those experiences was positive or negative. The data were analyzed both quantitatively and qualitatively to understand both the frequency and proportion of negative and positive experiences by experience type, as well as the themes and features that characterized these experiences. We then applied the resources vs. demands framework and identified instructional, cognitive, and motivational resources and demands that were common across the different experiences.

[2] An important distinction to make here is that the term "demand" should not be equated to "negative," and "resource" should not be equated to "positive." Rather, it is the alignment and balance of resources and demands that determine whether a situation is viewed as a threat and experienced more negatively or viewed as a challenge and experienced more positively.

## II. METHOD

The data were collected at a large, public mid-Atlantic university. Analysis was conducted on text transcripts derived from a series of semi-structured focus group sessions meant to understand students' experiences in introductory physics. Part of the goal of these sessions was to inform the development of a mindfulness training intervention for physics students. The training incorporated stressors and challenges reported by focus group participants as examples in order to tailor the intervention to the context of introductory physics. An overview of that project can be found on the Open Science Framework (https://osf.io/723rd) and we will not go into detail about it here; our focus is understanding students' experiences.

### A. Participants

Participants were 24 undergraduate students enrolled in at least one course of a two-term sequence of algebra-based introductory physics ($n$ = 13), or calculus-based physics ($n$ = 11). The mean age of the sample was 19 years ($SD$ = 1.25 years). The majority of the sample (91.7%) reported that they intended to complete a STEM major, and 50% of the sample were on an engineering track. The breakdown of gender and racial identities of the sample are provided in Table I[3].

[3] We collected the demographic data to give background and context to our sample, but our numbers in terms of individual participants were too small to make any meaningful generalizations about demographic groups.

TABLE I. Gender and racial identities of the study sample.

| | *N* | % of sample |
|---|---|---|
| **Racial and Ethnic Identification** (*Participants could select more than one*):[a] | | |
| African American/Black | 3 | 12.5% |
| Asian Indian | 1 | 4.2% |
| Asian/Pacific Islander | 3 | 12.5% |
| Hispanic | 1 | 4.2% |
| Non-Hispanic White | 16 | 66.7% |
| Not specified | 1 | 4.2% |
| *Students Who Selected Multiple Identities* | | |
| African American/Black/Non-Hispanic White | 1 | 4.2% |
| **Gender Identification** | | |
| Non-binary | 1 | 4.2% |
| Women | 18 | 75% |
| Men | 5 | 20% |

[a]Frequencies for racial and ethnic identification sum to greater than 24 (100%) of the sample because participants were able to select multiple race and ethnicity categories.

### B. Materials and Procedure

#### *1. Participant Screening and Recruitment*

Recruitment took place across two semesters. Research team members visited students in their classrooms to distribute recruitment flyers containing a brief description of the study and a link to the screening survey. An additional recruitment effort was made in which several instructors shared the recruitment flyer with students over email and/or Blackboard[4].

One hundred and thirty-five students completed the screening survey online using the Qualtrics survey delivery platform ($N_{\text{Semester1}} = 105$, $N_{\text{Semester2}} = 30$). Demographic characteristics collected on the screening survey were age, gender, and race and ethnicity. Respondents also reported information about their physics course, intended major, and English fluency. Four items were added in the second semester of the study, so data for these items were not captured for all participants. The four items measured year in school, prior physics experience, whether or not students were repeating their course from a previous semester, and if so, their grade from that semester.

In addition to the personal and course-related information, the screening survey collected responses to nine items measuring students' beliefs, feelings, and attitudes regarding their physics course (see TABLE II below). Five of the items were adapted from existing measures of beliefs and attitudes related to learning and motivation, such as social belonging [38], self-efficacy [39], fixed mindset [40], interest [38], and value [41]. The remaining four items captured students' feelings of stress, irritation, dread, and excitement regarding physics. Responses for each item were converted to a standardized z-score for responses to that item. FIG. 2 shows the mean item responses for focus group participants and non-participants.

[4] A learning management system similar to Canvas or Brightspace.

TABLE II. Screening survey items.

| Measure | Item | Response Options |
| --- | --- | --- |
| Interest [38] | Studying for physics is[a]… | 1 (Very boring) – 4 (Very interesting) [a] |
| Utility Value [38,41] | Studying for physics is[a]… | 1 (Not at all useful for accomplishing my long-term goals) – 4 (Very useful for accomplishing my long-term goals) [a] |
| Fixed Mindset [40] | To really excel in physics, a person needs to have a natural ability in physics. | 1 (Strongly disagree) – 6 (Strongly agree) |
| Self-Efficacy [39,41] | I feel I can do well in physics class[a]. | 1 (Strongly disagree) – 6 (Strongly agree) [a] |
| Excitement | Studying for physics excites me[a]. | 1 (Strongly disagree) – 6 (Strongly agree) [a] |
| Social Belonging [38] | I feel that I belong in my physics class[a]. | 1 (Strongly disagree) – 6 (Strongly agree) [a] |
| Irritation | Studying for physics makes me feel irritated. | 1 (Strongly disagree) – 6 (Strongly agree) |
| Stress | Studying for physics makes me feel… | 1 (Not at all stressed out) – 4 (Very stressed out) |
| Dread | I dread studying for physics. | 1 (Strongly disagree) – 6 (Strongly agree) |

[a]Reverse scored.

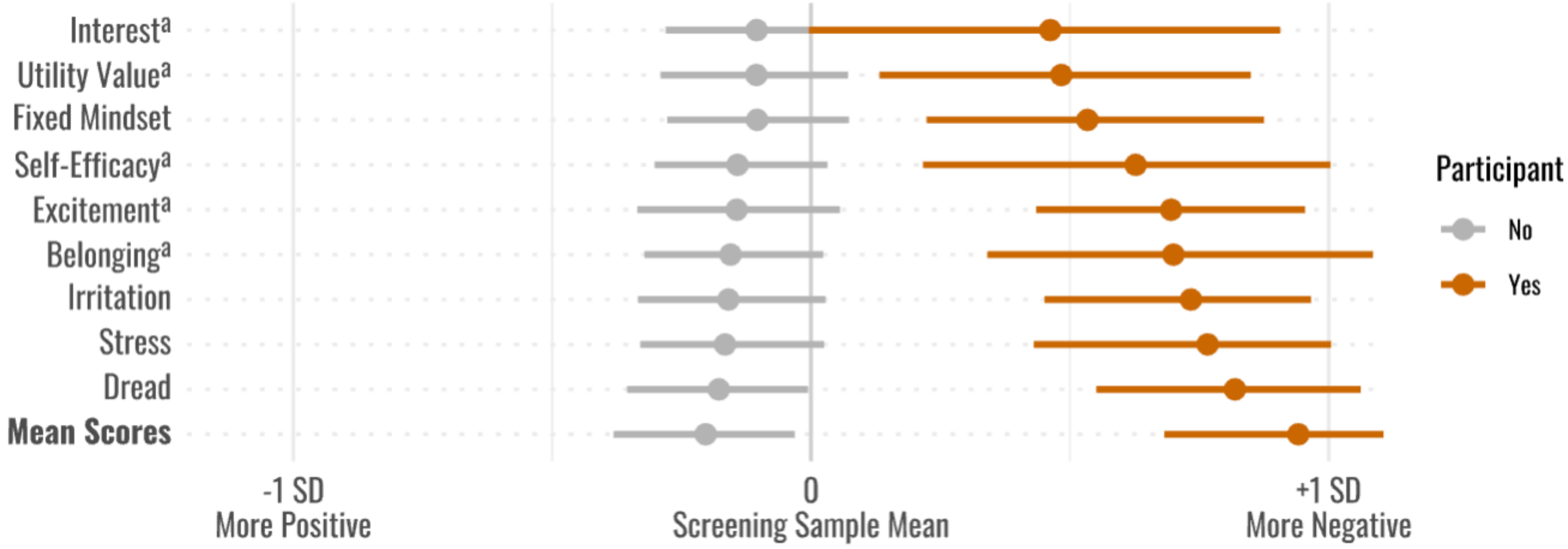

FIG. 2. [a]Reverse scored. Standardized scores for screening survey items and mean scores are shown for focus group participants compared to non-participants. The gray vertical line in the center of the plot represents the mean values on the screening survey items for all students who completed the survey (participants and non-participants). The point ranges on the horizontal lines show the means and bootstrapped 95% confidence intervals for the study participants (orange) and the remaining students (gray) for each measure.

A composite variable[5] indicating overall negative perceptions in physics was created by reversing the coding of all items except fixed mindset, irritation, stress, and dread (so that for all questions, a higher score indicated a more negative response) and averaging all 9 items. Only students with composite scores greater than 0 could be eligible for focus group participation. The composite measure had a Cronbach's alpha of 0.86 in the full screening sample ($N = 135$).

Sixty-nine students who were above 18 years of age, fluent in English, and had negative perceptions and experiences composite scores greater than zero or reported[6] that they were repeating their introductory course from a previous semester were eligible to participate ($N_{\text{Semester1}} = 52$, $N_{\text{Semester2}} = 17$). Eligible students were then recruited by phone to participate in a one-hour focus group session on campus. Participants were compensated $15 for attending the one-hour session.

### *2. Focus Group Session Protocol*

Twenty-four students participated in one of 13 focus group sessions during the second semester of the study. The number of participants interviewed per focus group session ranged from one to three: seven of the sessions were conducted in groups of two, four sessions were conducted with a single student, and two sessions were conducted with groups of three. Out of the final 24 participants, 19 completed the screening survey during the first semester and five were screened for eligibility and participated in the same semester (semester 2). Each session lasted one hour and two research team members were present for all sessions. Audio was recorded for the entire length of the session.

Before the sessions began, participants were presented with an overview of the study and verbally provided their consent to participate. After providing their consent, researchers began the recording and led the group in a framing activity (See Appendix A:).

---

[5] During the planning phase of the study, this composite variable was referred to as *risk of disengagement*. However, due to negative stigma associated with labeling students as at-risk, we decided to describe this variable as *negative perceptions and experiences in physics* to be more specific about what we were measuring.
[6] Repeating the course was only used as an eligibility criterion in the second semester of the study.

For the remainder of the session, researchers took turns prompting the discussion using the questions outlined in Table III. The questions were categorized into two sections. The first section was *open-ended* and consisted of four questions which asked students to generally describe their successes as well as aspects of the course they found challenging or stressful. The second, and more *specific*, section of the interview consisted of six questions which asked directly about experiences with instructors and peers, as well as students' goals and perceptions of their ability to do physics.

TABLE III. Semi-structured interview questions.

| | **Item** |
|---|---|
| **Open-Ended** | 1. What were the biggest challenges and successes during the semester? |
| | 2. What was the most stressful aspect of physics for you? Why? |
| | 3. What did you struggle with the most? What caused those struggles? |
| | 4. What were some of the successes that stand out? What caused those successes? |
| **Specific** | 5. How did you feel about the instructor and/or TAs in your introductory physics course? |
| | 6. Were there times you felt low confidence in your ability to do physics? |
| | 7. Were there times you felt confident in your ability to do physics? |
| | 8. Did you ever find yourself comparing yourself to your classmates? |
| | 9. What is your personal definition of "success" in physics learning? How does this relate to your short- and long-term goals? How did these goals change from the start and end of the semester? |
| | 10. Based on your experiences, what could have been done to improve your experience in your physics course? |

In sessions with more than one participant, each participant was given an opportunity to respond to each question, and students could refer to each other's responses. Researchers asked short follow-up questions when needed for clarification or to prompt elaboration. After one hour, researchers thanked participants for their efforts, stopped the recorder, and concluded the session.

### C. Analytic Process

#### *1. Audio Transcription and Text Segmentation*

Recorded audio from the interview and discussion portion of the focus group sessions was manually transcribed by research assistants. The transcripts were further reviewed for errors by a second team member and imported to NVivo qualitative data analysis software. Each transcript was then divided into segments by the first author for coding and analysis. The length of each text segment ranged from a single sentence (minimum word count = 68) to a short paragraph (maximum word count = 1,760). The average word count per segment was 563.39 (median = 519, mode = 483, SD = 297.6). The intention was for each segment to cover one experience. Therefore, sentences were kept together so long as the follow-up sentences were elaborations on the same point or descriptions of the same specific experience; responses were broken into multiple segments when a change of topic was noted or the student began describing another distinct experience. This was done to prevent the themes and views of less verbose participants from being underrepresented in the data [42]. Researcher dialogue was largely removed from analysis apart from a few special cases which included instances where the context of the preceding question was required for comprehension of the response or if the researcher asked a brief follow-up question for clarification. Segmentation for two of the 13 transcripts (15%) was reviewed by an additional team member to ensure consistency.

### *2. Experience Codes*

The experience codes were developed inductively, iteratively, and collaboratively with the research team [40, 41]. The final six[7] codes were: CLASSROOM AND COURSE STRUCTURES (CLASSROOM), CURRICULUM AND SPECIFIC LEARNING ACTIVITIES (LEARNING ACTIVITY), INSTRUCTOR, EXAMS, HELP-SEEKING, and PEERS (see TABLE IV for definitions). The goal was to identify the different types of experiences or situations students talked about, as well as the contexts in which they occurred, regardless of what those experiences were like (e.g., they could be positive or negative experiences).

Codes were applied to segments in full, regardless of whether the entire segment referenced the code. Additionally, codes were not mutually exclusive, meaning that multiple codes could be applied to a single segment. This was done because many segments were complex: comparing and contrasting activities across multiple contexts (e.g., working with peers in class compared to a study group) but phrased in a way that reducing or dividing the segment would result in fragments that were difficult to interpret. Applying multiple codes allowed us to use flexible filters and look at code co-occurrence (e.g., help-seeking with peers vs. help-seeking with instructors). To obtain the final coding data, all of the discourse was coded in full by two coders and all coding discrepancies were resolved through review and discussion until 100% agreement was reached. See Appendix B for a more detailed account of the coding procedure and reliability tests.

TABLE IV. Experience codes and definitions.

| **Code** | **Definition** |
|---|---|
| CLASSROOM AND COURSE STRUCTURES (CLASSROOM) | Experiences that occur in a classroom or recitation setting, can include broader experiences with course structure (e.g., flipped format) |
| CURRICULUM AND SPECIFIC LEARNING ACTIVITIES (LEARNING ACTIVITY) | Experiences relating to working through physics problems for practice, quizzes, completing assignments and/or homework, studying, clicker questions, or reading the book. Also includes talking about conceptual understanding of the material (course curriculum). |
| EXAMS | Experiences relating to exams, finals, and midterms, including preparation, test-taking, and reflections. |
| HELP-SEEKING | Experiences where the student has taken the initiative to seek help from another person(s) outside of class (e.g., attends office hours or tutoring, joins a study group, resource room, etc.) |
| INSTRUCTOR | Experiences relating to or involving the professor or TA(s). |
| PEERS | Experiences relating to or involving other students. |

---

[7] There were three additional codes at this level which are not described or defined here. Two of the codes, SUCCESS and SUGGESTION, encompassed abstract discourse that could not be evaluated as a course experience (i.e., definitions of success, suggestions for course improvement). The last code, SOCIAL COMPARISON, was a sub-category of PEERS, and therefore did not provide unique informational value. See Appendix B for a detailed account of the coding procedure and Appendix D for an analysis of text segments coded as SUCCESS, SUGGESTION, or SOCIAL COMPARISON.

FIG. 3 depicts an example of how the experience coding was applied to a single text segment. The student describes seeking help from their TAs after struggling with a homework assignment and contrasts the result of this experience with help they received from their peers. Therefore, the experience was coded to HELP-SEEKING, LEARNING ACTIVITY, PEERS, and INSTRUCTOR.

*I only went to visit the TAs once because I was struggling with the homework and I wanted to see if they could explain it, they just kinda went off and, you know, kinda like did it for me, and I didn't think that was really helping me. So, the TAs, the professor, trying to read the book and stuff, that wasn't anything compared to the help my friends gave me by showing me from a student's perspective*

HELP-SEEKING
LEARNING ACTIVITY
PEERS
INSTRUCTOR

FIG. 3. Complex example of experience category coding.

### *3. Positive and Negative Experience Coding (Affective Valence)*

After the experience codes were established, we wanted to be able to characterize the affective valence of these experiences; in other words, whether these experiences were perceived by students as positive or negative. A codebook was generated to define the characteristics of positive and negative experiences with respect to each experience code. Each experiential code was then coded as being either POSITIVE, NEGATIVE, BOTH (positive and negative) or NEUTRAL. Text segments labeled with multiple experience codes were coded as positive and/or negative for each associated experience.

Experiences were classified as NEGATIVE if the student explicitly mentioned it resulting in or involving emotions and/or feelings such as doubt, disappointment, frustration, hopelessness, or feeling bad. These included experiences described as unhelpful, even if the student's feelings were not explicitly stated, as well as barriers to accessing something that could have been helpful. For example, if a student described wanting to go to office hours, but not being able to go because of scheduling conflicts, that would be classified as negative.

Experiences were classified as POSITIVE if the student explicitly mentioned it resulting in or involving emotions and/or feelings such as happiness, excitement, interest, confidence, satisfaction, or feeling good. These included experiences described as helpful or beneficial, even if the students' feelings were not explicitly stated.

The NEUTRAL code was applied to experience codes that didn't have a positive or negative valence (e.g., mentioning an exam as a marker of time), and BOTH was applied when the experience had equal positive and negative qualities.

Referring again to the complex example in Fig. 3, the text segment was coded as NEGATIVE for INSTRUCTOR and LEARNING ACTIVITY because they struggled with the homework, reading the book was not fruitful, and they did not find their instructors helpful. The same text segment was coded as POSITIVE for PEERS and BOTH for HELP-SEEKING because they found the help they received from their friends to be beneficial, but their help-seeking experiences overall were both positive (peers) and negative (instructor).

### *4. Common Themes and Feature Identification*

After the positive and negative experience coding was complete, we identified common themes in the student experiences. Like the positive and negative coding, segments from each experience code were evaluated separately, such that segments tagged with multiple experience codes were evaluated for each associated experience type. Three members of the research team were assigned two experience codes each. On the first pass through the data, the researchers read through all segments coded as POSITIVE, NEGATIVE, and BOTH[8], and generated a summary of each segment in their own words to get an idea of the common themes. On the second pass, they assigned each segment to a theme and documented their themes in a separate text file, including specific examples as evidence for the theme as well as any features of the experience that were described in association with the theme. For example, feeling uncomfortable was a common theme among negative classroom experiences. Students attributed this discomfort to different features of the classroom, such as instructor behavior, class size, the reputation of the course, and the prior knowledge of their peers. Once each dataset had been reviewed by one researcher, they were exchanged and a new researcher reviewed each segment and indicated whether they thought the assigned theme was an appropriate characterization of participants' experiences. All discrepancies were resolved as a team.

## III. RESULTS

### A. Descriptive Overview

In total, 546 unique segments from the 24 participants were assigned to at least one of the experience codes and coded for affective valence (summary statistics are presented in TABLE V). The mean number of segments contributed by each participant was 22.75 (*SD* = 9.94). Each of these segments were coded with an average of 2.03 experience codes (*SD* = 1.00).

TABLE V. Summary statistics for experience and valence coding. The combined 1,107 total segments listed in the table is the product of 546 segments and 2.03 experience codes per segment.

| Code | Total Segments | Percent Positive | Percent Negative | Percent Both | Percent Neutral | Negative-to-Positive Ratio |
|---|---|---|---|---|---|---|
| Classroom and Course Structures | 156 | 10.9% | 69.2% | 11.5% | 8.3% | 6.35 |
| Instructor | 193 | 13.5% | 61.7% | 7.3% | 17.6% | 4.58 |
| Exam | 234 | 15.4% | 54.3% | 13.7% | 16.7% | 3.53 |
| Curriculum and Specific Learning Activities | 271 | 26.6% | 49.4% | 10.0% | 14.0% | 1.86 |
| Help Seeking | 89 | 30.3% | 47.2% | 10.1% | 12.4% | 1.56 |
| Peers | 164 | 34.1% | 32.9% | 11.6% | 21.3% | 0.96 |

The most frequently occurring codes were LEARNING ACTIVITY and EXAM, and the least frequent was HELP SEEKING. Frequency did not appear to be related to affective valence. For example, despite the fact that LEARNING ACTIVITY and EXAM were the most common codes, they were in the middle of the distribution with respect to the negative-to-positive ratio. The percent of segments within each experience affectively coded as BOTH or NEUTRAL was relatively stable across experience codes ($SD_{BOTH}$ = 2.1%, $SD_{NEUTRAL}$ = 4.5%) compared to the percent coded as either POSITIVE or NEGATIVE ($SD_{POSITIVE}$ = 9.8%, $SD_{NEGATIVE}$ = 12.6%). Therefore, the order of experience codes by percent negative is identical to the order of experience codes by negative-to-positive ratio.

[8] When identifying themes and features, we focused first on experiences coded as POSITIVE or NEGATIVE because they were the most straightforward and then reviewed those coded as BOTH to make sure there were no positive or negative themes that were qualitatively different than those already identified.

Figure 4 plots the relative proportions of POSITIVE, NEGATIVE, and BOTH codes for each experience code (excluding 42 segments affectively coded as NEUTRAL) broken out by whether students were responding to the open-ended (earlier) questions, specific (later) questions in the interview, or combining both question types (all). We analyzed answers to these question types separately to see if responses to the more specific questions were cuing a focus on certain kinds of experiences relative to the more open-ended questions. The affective valence proportions did not vary dramatically based on whether students were responding to the more open-ended (earlier) questions or more specific (later) questions. When considering which experience codes had greater proportions of reported negative and positive experiences, there were two differences between the open-ended vs. specific question types. The first is that whether the proportion of negative instructor experiences was greater than or less than the proportion of negative classroom experiences depended on question type. The same is true when comparing learning activity and help-seeking experiences. Despite these two differences, the proportional pattern of negative-both-positive in each interview section does not dramatically differ qualitatively from the overall pattern ("all" in Figure 4).

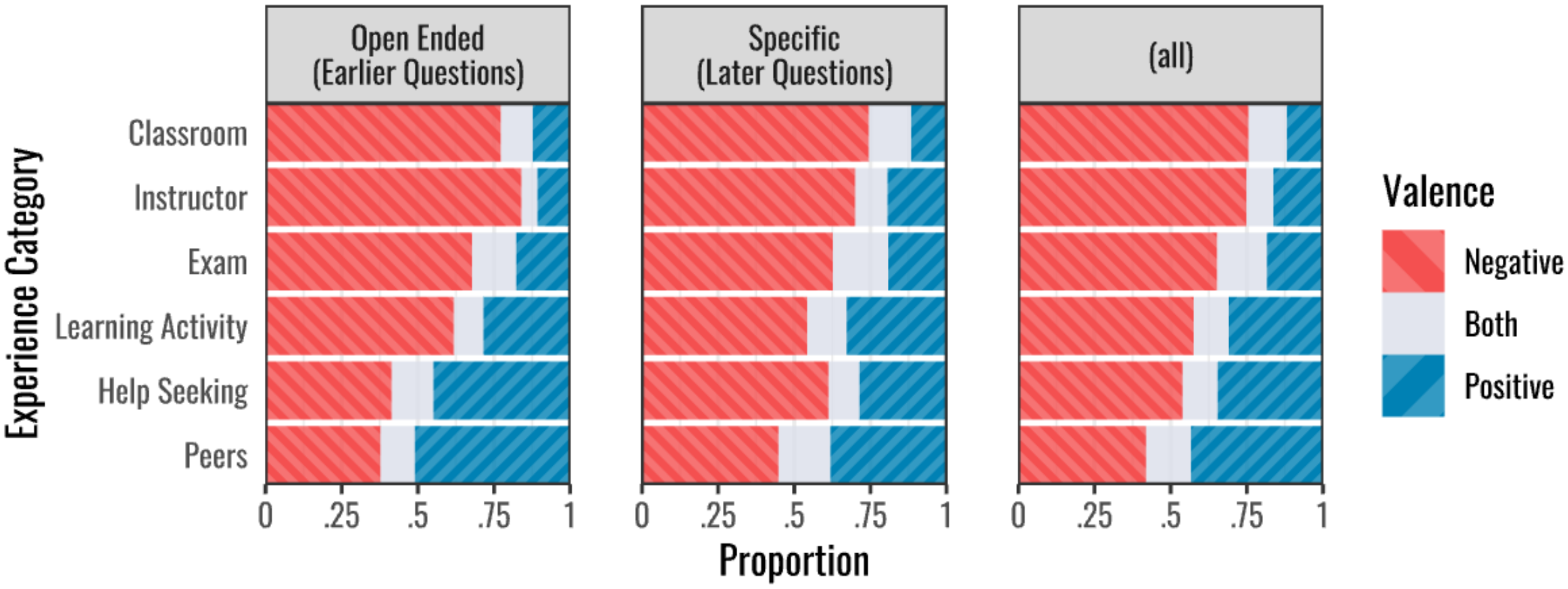

FIG. 4. Proportion of positive and negative segments by experience code, broken out by whether students were responding to the open-ended (earlier) questions, specific (later) questions in the interview, or considering both question types together (all). This figure includes the 504 unique segments coded as POSITIVE, NEGATIVE, or BOTH ($N_{open}$ = 218, $N_{specific}$=286).

TABLE V shows that the codes with the three highest negative to positive ratios and the three highest percentages of negative comments were CLASSROOM, INSTRUCTOR, and EXAM. The three lowest negative to positive ratios and the three lowest percentages of negative comments were PEERS, HELP-SEEKING, and LEARNING ACTIVITY.

### B. Summary of Themes and Features and Relation to Resources and Demands

To summarize the themes and features identified across the positive and negative experiences, we aggregated common elements and present them in TABLE VI (See Appendix C for a comprehensive list of themes and features identified across positive and negative experiences for each experience code). To organize themes and features and facilitate interpretation, they are presented along two dimensions. First, we organized them based on their relation to instruction, cognition, or motivation. Themes and features categorized as instruction include properties of lectures (e.g., level of abstraction, amount of review), classroom activities (e.g., demonstrations), course structure and/or environment (e.g., grading criteria, opportunities for interaction). Cognition includes themes relating to the students' perception of their own knowledge, understanding, and learning strategies. Lastly, themes and features categorized as motivation include students' reported beliefs, perceptions of themselves and others, and behaviors that they associated with their level of willingness or desire to engage in the course. The second categorical dimension is demands vs. resources. The resources vs. demands framework suggests that when demands outweigh resources a situation is more likely to be experienced as negative and vice versa. Consistent with this framework students often described extra or intense demands in their negative experiences and highlighted resources to cope with (often unstated) demands in the positive experiences. Therefore, we used the demands and resources distinction to help organize themes and features described in students' experiences. We note that demands and resources categorization is relative to specific situations and contexts.

TABLE VI. Summary of themes and features across experience codes

| | Demands | Resources |
|---|---|---|
| **Instruction** | Teaching style / course type (flipped vs. traditional) incompatible with learning style[a,b]<br>Not enough summarizing, reviewing, or explanation; rushing[a,b]<br>Too abstract[a,b,c]<br>Lack of alignment between class, text, homework, and exams[a,c,d] | Clear explanations (aligned with current level of understanding)[a,b,c,d,e,f]<br>Logic and/or rationale for problem-solving steps made explicit[b]<br>Real-world examples[a,b,d]<br>Engaging demonstrations[a,b]<br>Connections across concepts and domains[a,b,d] |
| | Lack of opportunities to ask questions[a] | Accessible office hours[a,b,e]<br>Opportunities for one-on-one interaction[b,e] |
| | Disorganized, lacking, or incorrect materials[a,b,d]<br>Lack of feedback[a,b,d]<br>Lack of clarity on grading criteria[a,b,c] | Low-stakes practice / practice materials[a,c,d,e] |
| **Cognition** | Assumed prior knowledge[a,b]<br>Confusing[a,d] | Prior knowledge / familiarity[a,c,d]<br>Effective study strategies[d,f] |
| **Motivation** | Intimidation and discomfort[a,b,e,f]<br>Anxiety, frustration, and stress[c]<br>Condescension and dismissiveness (from others)[a,b,e,f] | Belonging / solidarity / normalized struggle[a,f]<br>Acknowledgement (from others) of fear and anxieties; empathy[a,b]<br>Humor, enthusiasm, or encouragement (from others)[a,b,e] |
| | Fixed mindset; lack of physics identity[d,e]<br>Lack of utility value[a,d] | Evidence of improvement, learning, and mastery[a,c,d,f]<br>Good grade[c]<br>Adjusting goals/expectations[c]<br>Relevance to personal interests and goals[d] |
| | Negative social comparison, feeling judged or over-evaluated[a,c,e,f]<br>Time/resource constraints[c,d,e,f]<br>Diffusion of responsibility[a,f]<br>Falling behind[a,d] | Friendly competition; positive comparison[a,c,f]<br>Accountability[a,b,d] |

[a]Classroom and Course Structure
[b]Instructor
[c]Exam
[d]Curriculum and Specific Learning Activities
[e]Help-Seeking
[f]Peers

### C. Theme Co-occurrence among Positive and Negative Experiences

Next, we consider the distributions of negative and positive course experience codes across the text segments. TABLE VII shows the distribution of the 546 segments coded for affective valence. There were 350 total segments coded as negative for at least one experience code. For these negative experiences, the most commonly associated experience categories were CLASSROOM (108 segments), INSTRUCTOR (119 segments), EXAM (127 segments), and LEARNING ACTIVITY (134 segments). HELP-SEEKING (42 segments) and PEERS (54 segments) were less common. These more common experience codes appeared between 2 and 3.2 times more frequently than the less common codes, depending on which two codes are being compared. There were 159 total segments coded as positive for at least one experience code. For these positive experiences, the LEARNING ACTIVITY (72 segments) and PEERS (56 segments) experience codes were more common than the others, appearing between 1.6 and 4.2 times more frequently than other codes.

TABLE VII. Frequency and percentage of segments coded as positive or negative for one or multiple (co-occurring) experience codes

| | **Classroom** | **Instructor** | **Exam** | **Learning Activity** | **Help Seeking** | **Peers** | **All Codes** |
|---|---|---|---|---|---|---|---|
| **Negative Experiences** | | | | | | | |
| # segments with 1 negative code | 23 (21.3%) | 21 (17.6%) | 65 (51.2%) | 44 (32.8%) | 6 (14.3%) | 19 (35.2%) | 178 (50.9%) |
| # of segments w/multiple co-occurring negative codes | 85 (78.7%) | 98 (82.4%) | 62 (48.8%) | 90 (67.2%) | 36 (85.7%) | 35 (64.8%) | 172 (49.1%) |
| Total Segments | 108 | 119 | 127 | 134 | 42 | 54 | 350 |
| **Positive Experiences** | | | | | | | |
| # segments with 1 positive code | 2 (11.8%) | 10 (38.5%) | 21 (58.3%) | 37 (51.4%) | 7 (25.9%) | 27 (48.2%) | 104 (65.4%) |
| # of segments w/multiple co-occurring positive codes | 15 (88.2%) | 16 (61.5%) | 15 (41.7%) | 35 (48.6%) | 20 (74.1%) | 29 (51.8%) | 55 (34.6%) |
| Total Segments | 17 | 26 | 36 | 72 | 27 | 56 | 159 |

Second, when a negative/positive experience code appears in a segment, that code is more likely, on average, to appear in a multiply-coded negative/positive segment rather than on its own, indicating the intersecting nature of students' course experiences. TABLE VII provides the frequencies (and percentages) of how often each negative/positive experience code appears in segments (i) containing only that one negative and positive experience code or (ii) containing additional negative and positive code(s) (i.e., had multiple co-occurring negative and positive codes). On average, 71.2% of each negative experience code's occurrences were in segments containing multiple co-occurring negative codes, ranging from 48.8% of segments coded as EXAM to 85.7% of segments coded as HELP-SEEKING. On average, 60.9% of each positive experience code's occurrences were in segments containing multiple co-occurring positive codes, ranging from 41.7% of segments coded as EXAM to 88.2% of segments coded as CLASSROOM. Notably, the EXAM code has the smallest percentage of occurrences in multiply-coded segments for both positive and negative experiences, indicating that it is the least overlapping, most separable course experience. For negative experiences, this means that although EXAM is one of the most common experience codes, CLASSROOM, INSTRUCTOR, and LEARNING ACTIVITY have a larger number of co-occurrences with other experience codes within the same segment. Note also that the high percentage of experience code occurrences within multiple-coded segments does not mean that most segments have multiple co-occurring codes. Only 49.1% of segments have multiple co-occurring negative codes, and only 34.6% of segments have multiple co-occurring positive codes.

Mathematically, this reflects the fact that each individual multiply-coded segment is associated with two or more code occurrences.

Next, we consider which pairwise co-occurrences of valence codes are most common, to understand which types of intersecting negative or positive experiences are most common. FIG. 5 plots the frequency of pairwise code co-occurrences for negative and positive experiences. For both negative and positive experiences, the most common pairwise code co-occurrences are between the codes that appear most often within multiply-coded segments. For negative experiences (FIG. 5a), there are 306 total pairwise co-occurrences[9]. In considering which co-occurrences are notably large or small, we compare each cell in FIG. 5a to a criterion level of 20.4 co-occurrences: the value if the 306 pairwise co-occurrences were spread evenly across the 15 cells associated with unique code pairings. Only 6 of the 15 cells have greater than 20.4 co-occurrences. The three most common pairwise co-occurrences are the ones between CLASSROOM, INSTRUCTOR, and LEARNING ACTIVITY. These three pairwise co-occurrences account for 42.8% of all pairwise co-occurrences (131 out of 306 pairwise co-occurrences), showing how the intersection of classroom/course structure, the nature of the learning activity, and the instructors' actions are a major area of negative experiences for physics students. The high frequency of these pairwise co-occurrences are unsurprising given that these three negative experience codes generally occur most frequently in multiply-coded negative segments (TABLE VII): 98 multiply-coded negative segments with INSTRUCTOR, 90 multiply-coded negative segments with LEARNING ACTIVITY, 85 multiply-coded negative segments with CLASSROOM. The three other notably large cells indicate co-occurrences involving INSTRUCTOR and/or EXAM, the code with the fourth most multiply negative-coded segments: 32 EXAM – LEARNING ACTIVITY co-occurrences, 27 INSTRUCTOR – EXAM co-occurrences, and 22 INSTRUCTOR – HELP-SEEKING co-occurrences. Largely, these results reiterate that negative CLASSROOM, INSTRUCTOR, LEARNING ACTIVITY, and, to a lesser degree, EXAM codes most commonly occur in multiply-coded segments, as illustrated in TABLE VII.

For positive experiences, a similar qualitative analysis of FIG. 5b indicates that the most frequent intersections occur between LEARNING ACTIVITY, HELP-SEEKING, and PEERS. Given 98 pairwise co-occurrences, the average number of co-occurrences is 6.5, and 8 cells have a value greater than this. The three most common pairwise co-occurrences are the ones between LEARNING ACTIVITY, HELP-SEEKING, and PEERS. Pairwise co-occurrences between these three codes account for 36.7% of all pairwise co-occurrences (36 out of 98 pairwise co-occurrences). TABLE VII shows that LEARNING ACTIVITY and PEERS were the most frequent positive experience codes in general and within multiply positive-coded segments. Here, we find that the most common co-occurrences between these two codes are with each other and HELP-SEEKING. Of the remaining 5 notably frequent pairwise co-occurrences, 3 include LEARNING ACTIVITY (9 co-occurrences each with CLASSROOM, INSTRUCTOR, and EXAM) – another consequence of LEARNING ACTIVITY being the most frequent positive code overall and within multiply positive coded segments. Finally, 8 EXAM – PEERS co-occurrences and 7 INSTRUCTOR – HELP-SEEKING co-occurrence round out the notably frequent positive code co-occurrences.

[9] The number of pairwise code co-occurrences for each segment with N codes is ${}_{N}C_{2}$.

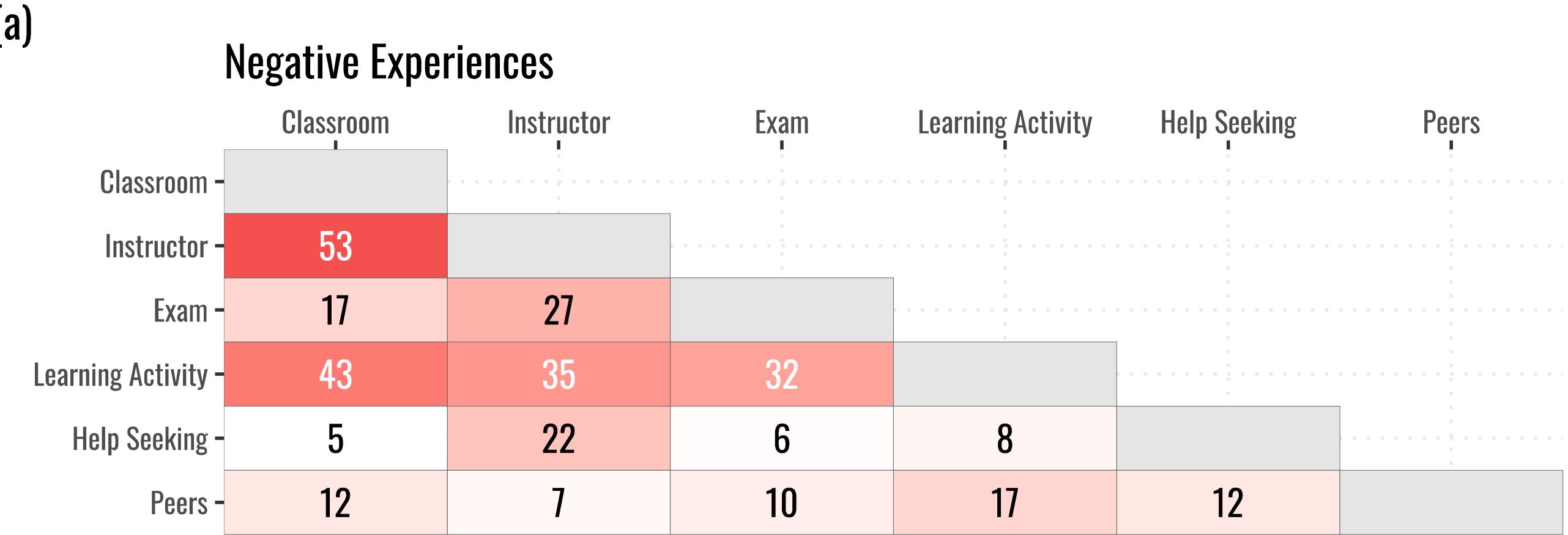

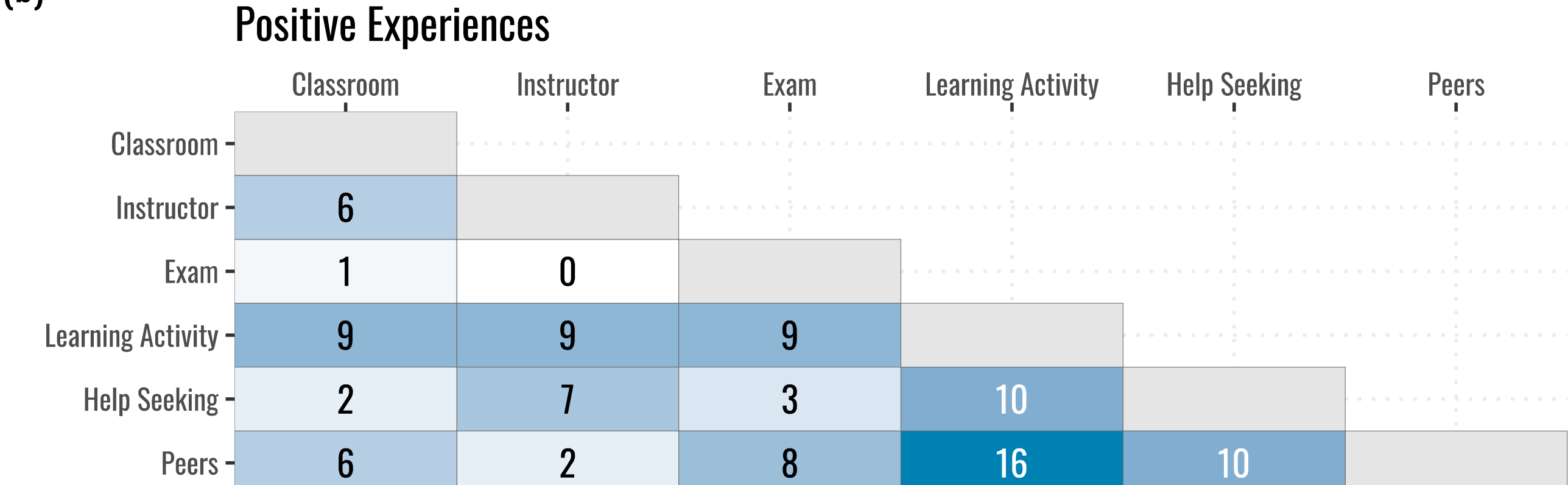

FIG. 5. Co-occurrence of negative (a) and positive (b) experiences by parent code. The numbers in the gray cells on the diagonals represent the number of segments within an experience category that did not overlap with any other code. The numbers in the lower-left triangle of the tables (cells shaded blue or red) represent the number of instances in which two experience categories overlapped in a single segment. As a reminder, each segment could be categorized with more than two categories, so some segments may be counted in more than one cell.

#### *1. Examples of Frequently Co-occurring Negative and Positive Experiences*

To help characterize the experiences students had that involved multiple experience codes, we present illustrative quotes[10] from the six most frequently co-occurring experience categories in FIG. 5. We chose to highlight these excerpts from students because they show the interconnected nature of student experience and the examples were carefully curated to collectively cover as wide a range of themes as possible. The examples are presented with their associated experience codes and themes to provide concrete instances of abstract categories.

*a) Negative experiences.* The three most frequently co-occurring combinations of experiences coded as negative were INSTRUCTOR & CLASSROOM, CLASSROOM & LEARNING ACTIVITY, and INSTRUCTOR & LEARNING ACTIVITY. The following student quotes contain one or more of these most frequent combinations. The experience codes and associated themes (see TABLE VI) are provided below each quote.

[10] Student quotes have been edited to remove filler words such as, "like" and "um," and for clarity. Gendered pronouns have also been replaced with gender-neutral ones to protect the identities of students and instructors. Omitted words are indicated with ellipses ("…"). Insertions are indicated with brackets (e.g., "she" is replaced by "[the instructor]").

*Yeah, and then with the professor I never asked a question like I would just like write it down and wait for another time… I remember I was… sitting on the end of a row once and… [the instructor] put up a question and I was like, I'm not doing it… I don't know where to start… I just kind of stared at it and [the instructor] came up to me and sat next to me and was… asking me how to do it and I… felt so uncomfortable just them being near me… I am… sweating right now thinking about it… I felt like I was gonna cry. I was like, I don't know… I was afraid to tell [them] I don't know how to do it and then [they] ended up… [they] looked at me and then stood up and walked away.*

Codes: CLASSROOM-NEGATIVE, INSTRUCTOR-NEGATIVE
Themes: Intimidation and discomfort

*It very much was like very detailed and very tricky to get into, and just like discerning what [the instructor] was actually saying because [they] would just assume that like... it was almost like [the instructor] was talking to us like we were like physics professors and not like actual students. And so, it was very hard to, like, get all of the terminology they were using and trying to understand everything all at once because [the instructor] wasn't breaking it down into a level like I would be able to understand.*

Codes: CLASSROOM-NEGATIVE, INSTRUCTOR-NEGATIVE
Themes: Confusing, assumed prior knowledge; teaching style

*But [the instructor] doesn't explain to the lowest member of the class, [they] explain it to the highest, so… if I'm in that 20% group that picked A and 80% picked C, [the instructor will] explain why C is right but [they] will not explain why A is wrong, and that was always really hard cause I'm just like, I'm wrong; I don't know why, so I have to go back and like relearn the material and at that point you're just like layering the same thing over and over so it just gets indistinguishable, what is correct and what isn't. That's always really difficult.*

Codes: CLASSROOM-NEGATIVE, INSTRUCTOR-NEGATIVE, LEARNING ACTIVITY-NEGATIVE
Themes: Confusing; Assumed prior knowledge; teaching style

*I feel like one of the hardest parts was feeling like our professor didn't really understand, like, at least my group of friends, when we would try to get help from [them], 'cause, like, so we expected from what [the instructor] told us, the first exam to be a lot different from how it ended up. Like, it was a lot harder, and [the instructor] said, like, "oh, time won't be a problem," and none of us finished, and all these things that were, like, you said one thing, but we experienced it very different, and we'd ask [them], like, I'm used to studying for bio and chem classes and this was way different. I felt like I never really figured out a good way to study for it, and we would go to [the instructor] and ask and [they] would be like, "oh, just look at the equations and figure out how they work," and for me that wasn't helpful. I don't know. It just like didn't click like it did for [the instructor]. And so, I couldn't figure out from anyone who was good at physics how to study for me in way that worked.*

Codes: EXAM-NEGATIVE, HELP-SEEKING-NEGATIVE, LEARNING ACTIVITY-NEGATIVE, PEERS-NEUTRAL, INSTRUCTOR-NEGATIVE
Themes: Lack of alignment between class, text, homework, and exams; teaching style / course type incompatible with learning style

*I went to [the instructor's] office hours like the first week, because I wanted to get ahead and stuff, and I remember one kid asked a question, and [the instructor] answered it, and the kid still didn't really understand, and [they were] like, "oh well, I don't really still get it," and [the kid] asked a follow up question, kind of like… [they were] like, questioning the professor, like [their] logic, kind of like, why that works. And [the instructor] answered [them] and was like,*

*"do I have a Ph.D. or do you?" And the kid was like, "I'm not trying to question your… knowledge, I just don't understand why that works." And like, after that, I never went back to see [the instructor]… why would I? No one was comfortable asking a question.*

Codes: INSTRUCTOR-NEGATIVE, HELP-SEEKING-NEGATIVE, PEERS-NEUTRAL
Themes: Intimidation and discomfort; condescension and dismissiveness

*I think the most stressful aspect was probably the lack of organization … because… I'm a person that in the beginning of the semester makes a spreadsheet for all my classes, like the grade book doesn't always take like weightings into account, so I always like to double check that I know what my grade is and, um, like we wouldn't get our exams back for a very long time, and we would ask, like, "hey, everyone has taken the exam it's been like a week so like anyone who had to make up probably took it. Can you upload a key, so we can have like an idea of how to study for the next exam or for our next quiz?," and [the instructor] was like, "oh, I didn't even make a key," and I was like, I would hope that you would make a key before you would make an exam to make sure that everything is correct. Like, even during an exam, [the instructor] would say something is wrong, and we would have to fix our booklets every single exam. So, just like the lack of organization was really stressful for me.*

Codes: CLASSROOM-NEGATIVE, EXAM-NEGATIVE, INSTRUCTOR-NEGATIVE
Themes: Disorganized, lacking, or incorrect materials; lack of feedback; teaching style

*b) Positive experiences.* The three most frequently co-occurring combinations of experience categories coded as positive were PEERS & LEARNING ACTIVITY, HELP-SEEKING & LEARNING ACTIVITY, and HELP-SEEKING & PEERS. The following student quotes contain one or more of these most frequent combinations. The experience codes and associated themes (see TABLE VI) are provided below each quote.

*I think that when we had, like, homeworks and I would have a group of friends and we would like meet to do them, I kind of felt more confident because I was with people who I didn't think would really judge me, and I was just like, "hey, I think this is the answer," and they would be like, "oh yeah, I actually think it is," I'd be like, "oh really, okay, let's keep going." I think those were the times that I was confident because I wasn't right in class, because in class, I kind of felt a little more pressure because there were so many people… where there will be people that are smarter than you, like a lot of people. Um I think, so when we had our take-home assignments, I felt confident.*

Codes: CLASSROOM-NEGATIVE, LEARNING ACTIVITY-POSITIVE, PEERS-BOTH, SOCIAL COMPARISON, SUCCESS
Themes: Belonging; evidence of improvement, learning, and mastery; discomfort; negative social comparison, feeling judged

*Yeah, um, before our second test I felt like I really understood our concepts, so I was helping other people out with them. Um, like I had a whole group of students I was working with and there were a few problems where no one was really sure, but I did something, and I had figured it out and then I had to explain it to everyone else and I felt good. Because if I can teach it then I can do the problem myself. And that's how I study now, like I get in a group, and we teach each other and like through teaching each other we get better at doing the problem…*

Codes: HELP-SEEKING-POSITIVE, LEARNING ACTIVITY-POSITIVE, PEERS-POSITIVE, EXAM-NEGATIVE
Themes: Prior knowledge / familiarity; evidence of improvement, learning, and mastery; effective study strategies

*I did like the fact that, um, we did problems, it was a flipped class so we did... more practice problems in class. I like the idea of clicker questions and that was helpful because it was an anonymous way of saying this is the answer I got and then [the instructor] would go over the problem. So that was, that was pretty good.*

Codes: CLASSROOM-POSITIVE, LEARNING ACTIVITY-POSITIVE, PEERS-POSITIVE
Themes: Low-stakes practice / practice material; clear explanations

*A lot of times I would see the TA that's in my recitation just like in the resource room because I just happened to go when [they] were in there, and [they] were a lot better one-on-one I think cause, like, learning physics one-on-one with someone helps so much more than in like a classroom setting just cause, like, its more comfortable and stuff. And um, so yeah, I would go to the resource room a lot to do like homework problems cause they would like walk you through it and then you get it.*

Codes: CLASSROOM-NEGATIVE, HELP-SEEKING-POSITIVE, LEARNING ACTIVITY-POSITIVE, INSTRUCTOR-POSITIVE
Themes: Discomfort; opportunities for one-on-one interaction; clear explanations

*I think it was like, yeah, like the second exam that I was talking about, where I scored higher than the rest of them… I tend to, on tests, skip questions that I don't know and come back to it and I think on that particular test, I had, when I revisited a particular question, I was able to figure it out. And it was probably because I was relaxing like, okay, need to like chill and just think about it and clear my mind and really really just grab from deep down where there's that information… I think it was because I had finished, like the open ended part and it was just like the multiple choice, there were just a few questions I was confused about. So, I was like, this is like the last thing I have to do and so it was fine for me to be like, okay just chill, and then approach the question.*

Codes: EXAM-POSITIVE, LEARNING ACTIVITY-POSITIVE, SUCCESS
Themes: Good grade; time / resource constraints

## IV. DISCUSSION

The resources and demands framework provides a theoretical structure for modeling physics students' experiences of challenge or threat. Considering these two independent dimensions of students' appraisals provides multiple possible pathways for promoting challenge and mitigating threat: increasing perceived resources, decreasing perceived demands, or doing both. Pragmatically, this work provides a language for addressing a common physics instructor question: "isn't making a course less threatening/more inclusive just lowering our rigorous learning standards?" While high learning demands are a necessary part of learning physics, researchers can seek to characterize whether proposed interventions increase resources and/or lower extraneous demands while holding the learning demanded of students at the same level of rigor. While we expect that many experienced instructors and researchers are intuitively attuned to the separate dimensions of resources and demands, articulation and investigation of this framework formalizes these ideas and can make them the explicit topic of instructional decision-making and further research.

Below we discuss the answers and implications to the following question: How can we use information about the landscape of student experience—information about breadth, frequency, and positive/negative associations of different types of experiences—along with the resources vs. demands framework to further our understanding of the class ecology and opportunities for change? We also propose a distinction between extraneous and necessary demands and suggest strategies for identifying and addressing them. Finally, we describe the limitations and next steps of this work and some concluding thoughts.

### A. Setting Priorities and Targets for Improvement

We analyzed student discourse to understand the types of experiences they had in introductory physics and further classified these experiences as positive or negative to contextualize them within the resources vs. demands framework. According to the theory, students are more likely to have positive experiences in a state of challenge (resources > demands) and negative experiences when they are in a state of threat (demands > resources). Therefore, the negative-to-positive ratios for experience codes can serve as an indicator for which experiences are more likely viewed as challenging or threatening for students, and—along with information about code frequency—help instructors set priorities for *areas of intervention*. The themes and features associated with those experiences bear insight into what resources and demands are relevant for different types of experiences, and therefore, provide clues as to what instructors *can intervene upon* to effectively shift the balance of resources and demands in specific contexts.

The three codes that had the largest negative to positive ratio scores were classroom, instructor, and exams. This observation suggests that these are areas in the classroom ecology where students are reporting greater threat and may require the most change. Much work in disciplinary based physics education and cognitive science has focused on creating interventions that impact the student learning, motivation, and performance outcomes through changes in *learning activities* [see [44–47] for reviews]. There appears to be less work investigating instructor interactions and exams as targets of intervention [47] relative to learning activities though these categories have higher negative-to-positive experience ratios[11]. Exam and instructor were also the two most frequently occurring codes, which suggests that interventions targeting these experiences could have greater impact. Taking *exam* as an example, instructors could try to minimize grading ambiguities and help students navigate test anxiety (reduce demands) or provide opportunities for students to improve and demonstrate mastery (increase resources), such as quiz corrections [48].

Another way to prioritize areas of intervention is to select low-frequency experiences that students tend to experience as more positive and less threatening and increase them. In our analysis, the two most positive (or least negative) experiences were help-seeking and peer engagement. They were also the first and third lowest frequency experiences, respectively. Prior work has shown that help-seeking behaviors are often associated with positive learning and motivational outcomes [see Refs. [49–51] for reviews] and the current results are consistent with this work in that this category is associated with features of encouragement, accessibility to one-on-one instruction, and clear explanations. Our results are also consistent with much prior work that has found that peer collaboration can be productive for classroom motivation and learning outcomes [see Refs. [52–54] for reviews]. The current work shows that this category was associated with features of belonging, clear explanations, effective learning strategies, mastery, and positive comparison. Though there are times that peer work can lead to unproductive and unhelpful interactions as well such as when they are associated with fear of judgement, negative social comparison, and diffusion of responsibility. An instructor attempting to incorporate more group activities can use this information and maximize the positive impact of their efforts by being mindful of these potential demands.

Finally, our analysis of co-occurring codes shows the multifaceted nature of student experiences and suggests that interventions targeting a specific student experience may impact perception of multiple features of the class. For example, interventions that focus on learning activities may have potential for decreasing multiple dimensions of negative experience related to instructor and classroom but also the potential to increase positive aspects of peers and help-seeking.

### B. Extraneous vs. Necessary Demands

When devising strategies to address student demands, it may be helpful to think about their nature. For example, some of the demands we identified can be considered *extraneous*, meaning that they are not directly related to learning or logistic constraints (e.g., clarity on grading criteria). *Necessary* demands are either *unavoidable*, such as individual differences (i.e., students' level of prior knowledge) or practical limitations (i.e., time constraints, class size), or *fundamental to learning*: having to do with the substance of physics knowledge and related skills, and the inherent difficulty of their acquisition [55,56]. These demands are directly related to course objectives and the learning goals set for students.

[11] However, see efforts by [82] and [83] for work on instructor interventions and [84,85] for work on assessment interventions.

Our interpretation is consistent with—but not identical to—two of the main components[12] of cognitive load theory: intrinsic and extraneous cognitive load [57–59]. Cognitive load can be thought of as the amount of mental effort required by working memory to process information during learning, where intrinsic load is the product of information complexity and the learner's (lack of) expertise, and extraneous load is any additional load imposed by instructional choices or environmental distractions. Our conceptualization of necessary demands completely encapsulates the intrinsic cognitive load imposed on students to learn physics, and there is much overlap between extraneous demands and extraneous cognitive load. Additionally, the idea of balancing cognitive load and working memory capacity is analogous to the idea of balancing demands and resources. However, the key distinction between our framework and cognitive load theory is that our conceptualization of demands and resources is more holistic and expansive, encapsulating motivational (social, emotional) components of student experience in addition to cognitive and instructional.

According to the resources vs. demands framework, a challenging situation is not absent of demands; rather it is one in which students have adequate resources that are also in alignment with those demands. Our premise is that a challenging physics course need not be threatening and should be challenging for the purpose of learning, and not due to an excess of extraneous demands. This means reducing extraneous demands and supporting students' resources to meet necessary demands.

Features associated with students' positive experiences with instructors, classrooms, and structures provide a basis for identifying which resources could be most effective for addressing necessary demands. For example, students had positive experiences when explanations provided to them were aligned with their level of understanding, when the logic and rationale for problem-solving steps were made explicit, when instructors made efforts to connect to real-world examples and across domains, and when opportunities for one-on-one interactions were available, accessible, and actively encouraged. These observations are consistent with the learning sciences literature showing benefits of instructional explanations [60–62], self-explanations [63,64], use of real-world examples [65,66], and tutoring [67–70] on learning and performance. These resources may be especially helpful for those students who felt that instructors and/or lectures assumed prior knowledge they did not possess. Benefits for these students could be compounded by using pretesting and low-stakes formative assessments to identify and monitor knowledge gaps and then use this information to tailor scaffolding and support to student needs [71].

We understand that the nature of a specific demand may vary based on context, especially the difference between those considered extraneous and unavoidable. For example, a time constraint may be unavoidable in one situation but extraneous in others. Class size may be determined by an institution's available facilities and enrollment numbers, which are outside of the control of an instructor. Therefore, we propose these distinctions as a useful guideline for thinking intentionally about student demands and provide examples of demand types but leave it up to researchers and practitioners to determine whether demands are extraneous, unavoidable, or fundamental to learning within their learning context.

### C. Limitations and Future Directions

We see two main limitations of the current work. The first is with respect to aspects of internal validity and the second relates to generalizability of the findings. Our primary research goal was to describe the experiences and course ecologies of students with negative perceptions towards physics and their physics course. One factor that likely influenced our participants' perception of their experiences is that they were reflecting on their previous semester experiences. This delay in time may have had an impact on what features of the class were recalled and could have been influenced by students' ultimate performance in the class. We expect that the reflections after the semester would likely share both similarities and differences to how those experiences would be perceived during the semester. Consistent with theories of remembering, we view memory recall as a constructive process that is influenced by many factors including time since the experience [72], context of recall [73], prompting questions [74], other responses [75], etc. Thus, there may be experiences that are more likely to be reported during the semester than after that our methods do not capture.

---

[12] Cognitive load theory also identifies germane cognitive load as effort required for schema construction. See [58] for more detailed definitions and distinctions.

Yet, there are also some potential advantages to conducting the interviews after the semester. The passage of time allowed students the space to make sense of their experience and contextualize it within their broader academic journey before trying to articulate it. Additionally, memory may have served as a sort of filter which selected for the most emotionally salient experiences and painted a picture of the types of situations that are most likely to stick with students after the class is over [76]. In addition, our participants were screened and selected for recruitment while the course was still in progress, so their outcome in the course could not have affected their perceptions at that time. Another advantage of waiting until after the semester is that participants may feel freer to share their experiences because the class is over.

With respect to generalizability, our explicit goal was to understand the experiences of a specific subsample of students—those who reported greater negative perceptions relative to their peers. Future work could further investigate whether these observations are also present for those who report more positive perceptions of the class. We propose that the demands and resources framework would be a helpful tool to describe and understand student experiences, even for this different subgroup.

Finally, our goal was to provide rich, qualitative descriptions of student experiences. While our sampling of 24 students was good for revealing the range of possible resources and, it is not a large enough sample to make claims about characteristic experiences in different lecture sections (i.e., flipped format instruction), especially given that these students were recruited from multiple lecture sections, with different content, pedagogies, and instructors. Furthermore, students' qualitative reports were complex and even contradictory at times. For example, some students reported that a surplus of graded assignments made them feel like they were constantly being evaluated, while others wanted more graded assignments in between exams to increase accountability; some students enjoyed the flexibility afforded by their flipped format course, while others felt isolated and found it difficult to carve out the time to stay on top of asynchronous lectures.

Therefore, we caution against drawing simple conclusions about instructional features that students experienced positively and negatively as there is no single approach or instructional feature that will be ideal for all students. However, our results can provide insights into how and why certain features of physics learning experiences create resources and demands for students. Researchers can use these insights to test whether providing certain resources offsets the demands associated with a particular instructional feature. For example, normalizing mistakes, highlighting the learning benefits of formative assessments, or giving students opportunities to increase their score with effort, may help students receive the accountability benefits of graded assignments without feeling like they are being over-evaluated.

### D. Conclusion

This work sheds some light on the types of experiences students have in large, demanding university courses, and how features of the courses are perceived by students. We provided a framework for thinking about how to relate aspects of instruction, cognition, and motivation to student experiences. For instructors with some intuitive understanding of these issues, the framework provides a way to formally evaluate and discuss their instruction. For others, the resources and demands framework can introduce a new reflective lens for considering teaching practice. A key idea to take away is that resources and demands will vary depending on the context of a situation. Researchers and instructors can design and evaluate instructional features on the basis of the types of demands they create for students and whether they contribute to the resources students have to meet those demands. We propose that instructional features that reduce demands not directly related to learning and intentionally support student resources aligned with their demands will be most effective for creating positive learning experiences for students. We hope that future research can use this framework as a basis to test course transformations and interventions.

Finally, we suggest the possibility that often, when instructional choices place extraneous demands on students, it is because those choices are low-cost for instructors. When instructors make choices that prioritize convenience and ease, the cost of these choices may be passed on to students. For example, it is easy to list office hours in the syllabus, but it takes more effort to encourage students to attend or think about ways to make that office hours experience less demanding and more productive once they are there. We hope this work can support the redesign of physics courses so that more students can experience learning physics as a challenge and not a threat [77,78].

## ACKNOWLEDGMENTS

The authors would like to thank Gillian Murphy and Brandon Baun for their help with data collection. We extend additional thanks to Sara Jahanian, Kristen Hinshaw, Keely Lombardi, Brennan Paik, and Aliah Zewail for their help with transcription and coding. This work was supported in part by grants from the John Templeton Foundation, Award No. 5429; the National Science Foundation, Award Nos. DUE-1524575 and DUE-2100040; and the James S. McDonnell Foundation, Award No. 220020483. Additional support was provided by the University of Pittsburgh through the Provost Personalized Learning Award, the Discipline-based Science Education Research Center, and the Learning Research and Development Center.

## APPENDIX A: JOURNEY MAPS

The purpose of this icebreaker activity was to help students reflect on their experiences in their physics class [79] before the interview. Researchers adhered poster-sized pieces of paper to a wall in a portrait orientation and drew an "S"-shaped pathway diagonally from the upper left corner to the lower right on each sheet. Students were provided with a pack of colored markers and asked to annotate the pathway to represent any challenges, successes, or milestones they experienced along their journey through physics class. They were instructed to use any combination of words and images to depict their experiences.

## APPENDIX B: CODE DEVELOPMENT AND RELIABILITY TESTING PROCEDURE

### 1. Experience Codes

After reading the transcripts and experimenting with various coding schemes, the first author proposed a set of 11 codes and drafted corresponding definitions, inclusion, and exclusion criteria. Five team members then independently coded two transcripts based on the codebook. Percent agreement and unweighted Cohen's Kappa values were calculated for each pair of coders, and the mean values for all pairs were calculated for each code. Coding discrepancies were discussed by the larger team and the codebook was revised based on these discussions. This process was repeated three times on unique transcript samples (approximately 15% of total transcripts for each iteration) until the mean percent agreement reached 91% and the mean overall Kappa value reached .66. Because the first three iterations saw acceptable reliability for seven codes (Kappa values between .66 and .83), for the fourth and final iteration on new content, the team only coded for the four codes with the lowest reliability (Kappa values between .4 and .54). Following this iteration, two codes (*self-evaluation* and *motivation or goals*) were dropped because coding reliability failed to increase. The Kappa values for the remaining two low-reliability codes (*classroom* and *help-seeking*) increased by .19 and .16 points respectively and were retained, resulting in a final set of nine codes with the mean Kappa ranging from .58 to .82 ($M$ = .72, $SD$ = .07) across the remaining set of codes and coders.

Once the experience codes were established, all thirteen focus group transcripts[13] were then coded in their entirety by two coders. The overall unweighted Kappa for the full text coding was .74 and the mean percent agreement was 93%. All coding discrepancies were resolved through review and discussion by the two coders until 100% agreement was reached.

### 2. Positive and Negative Experience Coding (Affective Valence)

To categorize experiences as *positive*, *negative*, *both* (positive and negative), or *neutral*, a separate spreadsheet was created for each of the six experience codes. Each text segment constituted one row in the spreadsheet. Segments could appear in multiple spreadsheets if they were coded with multiple experience codes (e.g., CLASSROOM and INSTRUCTOR).

Two research assistants then independently coded the first half of the segments in each spreadsheet. Then, coding discrepancies were reviewed by a third coder (the first author). For each discrepancy, the third coder made a judgment and provided a written explanation in the notes. The first two coders reviewed these notes and had the opportunity to ask for clarification. Any issues were resolved through group discussion and if needed, minor adjustments to the wording of the codebook were made. After this checkpoint, the first two coders finished the coding the remainder of the segments. Weighted Kappas for the valence codes ranged from .61 to .68 and the percent agreement ranged from 71.5% to 80.8%. All discrepancies were resolved by the group.

[13] As a reminder, focus groups were conducted with groups of one, two, and three participants, so 13 transcripts contained verbal data from all 24 participants.

## APPENDIX C: THEMES AND FEATURES CODING RESULTS

The following sections describe common themes and features of students' negative and positive experiences[14] by experience category, starting with the highest negative-to-positive ratio and ending with the lowest. Within each category, the themes are listed in approximate order of frequency, starting with the most commonly occurring theme to the least common.

### 1. Classroom and Course Structures: Themes and Features

#### *a. Negative Experiences*

Students talked most frequently about feeling uncomfortable or intimidated in the classroom setting. This was sometimes due to specific behaviors attributed to the professor, but also attributed to aspects of the environment, such as large class size or the reputation of the course as being a 'weed-out' class. Students talked about lectures or recitation sessions feeling rushed or disorganized, and materials being confusing. Students who were in a flipped-format course talked about several challenges they faced with this format. These challenges included the inability to ask questions during pre-recorded asynchronous lectures, difficulty keeping up with the lecture videos, and a decrease in motivation to come to class after falling behind. Students experienced frustration about feeling like they weren't learning in the classroom setting. Sometimes this was related to the disorganization mentioned above but was also related to in-class demonstrations and problem-solving activities. Some students expressed that in-class activities were too simple or general and did not prepare them for exams or homework. Others said that the problems and/or activities were not relevant to their academic interests and not engaging enough to hold their attention. Students reported negative experiences during in-class group work when they felt like they didn't have enough instructional support, when they couldn't or didn't feel comfortable contributing to the group, or when they felt like they weren't learning because it was too easy to defer to other students for answers. Finally, students talked about communication barriers in the classroom. In some instances, this was a language barrier, in others, students felt that the lecture assumed prior knowledge they did not possess and that information was not conveyed at a level they could understand.

#### *b. Positive Experiences*

Most commonly, students talked about specific aspects of instruction they experienced as positive. In terms of in-class experiences, these included relevant and interesting demonstrations, well-organized lectures, summaries at the beginning of class, time for additional review and/or explanation for difficult topics, explicit connections between different content topics, and humor. More broadly with regard to course structure, these included accessible scheduling of office hours (e.g., close to class time) and graded homeworks, to increase accountability. Students who had positive experiences with the flipped-course format reported that they liked being able to set their own schedule and pace. Group work in the classroom was viewed in a positive light when other students could explain ideas in a manner that was more accessible to them or when they viewed comparison and/or competition (e.g., answering a clicker question and seeing the distribution of responses) as friendly and motivating. Students also reported feeling less alone when they struggled in class compared to on their own. For example, if they felt lost or confused on a topic, it was positive if they knew that others were feeling the same. Finally, students reported having positive experiences in the classroom when they felt like they were able to follow along and successfully solve the in-class examples.

---

[14] Positive and negative themes and features from segments that were coded as *both* positive and negative with respect to an experience code are listed separately.

### 2. Instructor: Themes and Features

#### *a. Negative Experiences*

Most commonly, students talked about negative experiences with their instructors' teaching and/or lecturing style. In some cases, there were general comments about how the teaching style was ineffective or incompatible with how they learn. A few more specific features of teaching style that were viewed negatively were that the instructor's style was too abstract and didn't incorporate (or connect to) real-world or specific examples, and that there was not enough time spent on reviewing and/or summarizing content before jumping into something new. Some students felt that the instructor moved too quickly through topics and valued deadlines over students' understanding and would cater to select students in the class with more prior knowledge as opposed to those with less. Next, students talked about instructors failing to adequately explain concepts or problems when students asked questions. Specifically, they felt that instructors had a hard time explaining their own problem-solving steps and the logic they used to get from one step to the next. They found it unhelpful when instructors simply gave them answers or directed them to a slide and/or section in the book instead of walking through a problem from the beginning. It was also viewed as unhelpful when instructors would go through the steps of problems too quickly without explaining the rationale for each step.

Students also talked about how aspects of their instructor's interactions contributed to negative experiences. In these experiences, students described instructors as being unapproachable or condescending, non-empathetic, dismissive, or showing a lack of enthusiasm about teaching. Additional themes included lack of support or resources provided by the instructor, disorganized or confusing class materials (e.g., slides had illegible handwriting or mistakes in the calculations), lack of prompt feedback on assignments, and lack of alignment between content covered in lecture with exams. Finally, a few students talked about dissatisfaction with the grading scheme set by the instructor. Specific complaints were lack of clarity in grading criteria, not enough graded assignments between exams to promote accountability, and one student mentioned that underweighting exams in the final grade made it seem like their studying was pointless.

#### *b. Positive Experiences*

When students spoke positively about their professors and TAs, they generally talked about aspects of teaching style and interactions. With respect to teaching style, students appreciated instructors who were able to provide clear explanations at a level that was easily understood by students, made connections to other concepts and domains, conducted engaging demonstrations, displayed an enthusiasm for the material, and incorporated humor into the lectures. Students also appreciated when instructors made efforts to make themselves available for office hours and one-on-one engagement and actively encouraged students to access these resources. With respect to interactions, students felt positively when their instructors acknowledged their fears and anxieties and provided reassurance and empathy.

### 3. Exam: Themes and Features

#### *a. Negative Experiences*

One of the most common themes among negative experiences with exams was that they didn't match students' expectations, either because students felt like the exam questions were too different from the problems covered in class or on homeworks, because they were simply too difficult (e.g., problems were too complex or conceptual) or because the professor changed the test format unexpectedly (e.g., multiple choice final after short answer midterms). Dissatisfaction with grades or grading scheme was also very common. Additionally, students expressed frustration and hopelessness when they felt like they made great efforts to study but could not see their labor reflected in their performance and reported feeling stressed and anxious before exams. Students also reported experiencing negative emotions when they were being outperformed by their peers. Finally, some students felt like they were being tested too much, while others felt like they were not tested enough, although the latter sentiment was less common.

Students who reported too much testing either felt like they were being constantly evaluated within their physics class or that their physics exams would too often line up with exams in their other courses and they had to choose which exam to study for. Students who wanted more testing felt that more low-stakes quizzes between exams would have helped them regulate their studying through the semester.

#### *b. Positive Experiences*

The three most common themes in the students' positive exam experiences were grades, prior experience, and adjusting expectations. With respect to grades, students were happy when they received a score that met or exceeded their performance goals. However, grades were not the only feature of exams that students talked about positively. Having prior experience with physics material (e.g., from high school) or familiarity with the exam structure and well-developed study strategies from earlier in the semester helped students feel confident and prepared. Another thing that helped students feel better about exams was adjusting their expectations about performance. For example, students who had very high absolute standards for academic performance shifted to performance standards that were more relative, such as personal improvement or contextualizing their performance in comparison to their peers, and this helped them feel better about exam performance. Students also reported feelings of satisfaction when they studied hard and felt like this effort paid off during the exam, or when they got to the exam and felt like they really had a grasp on the material and knew how to navigate the problems. It is important to note that feeling like study efforts were rewarded and having a grasp on the material likely went hand-in-hand with better performance, but the final performance outcome was not the only aspect of the exam experience students described as positive. Finally, students reported positive experiences when they had access to a resource—such as a TA or a peer—who was able to give them some advice about what to expect on the exam, or when the professor modified the format of the exam in a way that was more familiar to them (e.g., moving to multiple choice from open-ended questions).

### 4. Curriculum and Specific Learning Activities: Themes and Features

#### *a. Negative Experiences*

The most common themes among the negative experiences with the course curriculum and specific learning activities were the inability to find effective study strategies, lack of improvement, and feeling like study efforts were futile. When studying, students felt like they had trouble developing more than a shallow understanding of the material and difficulty making connections between concepts, which was frustrating and made them feel like they didn't have a "mind" for physics. Students talked about feeling confused because the problem explanations provided to them didn't make sense or they felt like there wasn't enough time to grasp the full breadth of topics covered. They also felt like the problems they were given in class or as homework did not adequately prepare them for exams, and that there were not enough useful practice materials provided. Procrastination and falling behind on assignments or asynchronous activities were also mentioned as negatively impacting learning. Students generally had negative experiences when they felt underprepared for the class because of a lack of prior knowledge or experience (i.e., from high school or in related subjects, like calculus). Students who had negative experiences during problem solving reported feeling like they had a lot of trouble getting started with problems and moving forward when they got stuck. Finally, some students mentioned the workload being simply too heavy or opportunities for study (e.g., peer study groups) being inaccessible.

#### *b. Positive Experiences*

Students described feeling excited, confident, and successful when things "clicked" and they were able to move forward with a problem or reach a higher level of understanding. Students also reported increased interest and value when they could connect physics concepts and their personal interests and goals. These connections were facilitated by problems and examples that were relevant to other domains (e.g., human physiology). Some students reported this to be true regardless of whether they were able to get the entire problem correct in the end. It was important for students to feel like their effort and engagement was resulting in noticeable improvement. Working on practice problems that were similar to exam problems was a particularly helpful activity, although students also found external resources such as YouTube or Khan Academy helpful as well. Having required—as opposed to optional—homework was useful for ensuring accountability (optional homework was too easy to not complete) and having multiple opportunities to get the homework problems correct helped relieve performance pressure. Some positive features of classroom learning activities (e.g., clicker questions) were that they were a low-stakes and anonymous way to get practice, and students appreciated the immediate feedback and explanations.

### 5. Help-Seeking: Themes and Features

#### *a. Negative Experiences*

When seeking help from instructors, students reported feeling put off due to the instructor's teaching style being condescending or dismissive. Students also talked about help being inaccessible to them. Barriers to accessibility of help with respect to office hours were scheduling and/or logistic barriers (e.g., office hours occurred at times when the students had other obligations), lack of advertising by the instructor about office hours, and scarcity of time and/or attention due to too many students seeking help at the same time. Scheduling and logistic barriers were also a factor for seeking help from peers, as was not knowing anyone in the class. Negative social comparison was a major factor when seeking help from peers. When students worked with their peers, asking for help made them feel like they were slower or less intelligent in comparison to others. Another theme was a lack of structure provided for help-seeking activities such as office hours. For example, some students felt uncomfortable in office hours because they knew they needed help but didn't know what questions to ask. These students felt like a structured activity, such as a worksheet, would have helped them generate specific questions and made the experience less intimidating. Students also mentioned avoiding office hours because it felt like a blow to their pride to seek help.

#### *b. Positive Experiences*

Students who had positive experiences seeking help from their peers reported that other students were able to explain concepts and problem-solving approaches at a level that they found accessible. Students also talked about one-on-one attention and having the time to work through a set of problems in depth during office hours or tutoring sessions to be beneficial. These types of experiences helped increase confidence, bolster feelings of mastery, and relieve stress and anxiety. Finally, students had positive experiences when help-seeking opportunities were made accessible. For example, having office hours scheduled directly after class made it easy for a student to attend without having to make a special trip on a different day, and this boosted their confidence.

### 6. Peers: Themes and Features

#### *a. Negative Experiences*

Negative social comparison was a common theme in negative experiences with peers. Students described feeling bad when they felt that coursework and physics learning came easier or faster for peers, when they felt they had less prior knowledge compared to their peers, or when they scored lower than the class average. Students also talked about feeling isolated and alone when they were not able to connect with their peers either due to scheduling and/or logistics or intimidation and fear of judgment. Finally, experiences deferring to peers for answers and avoiding working through a problem for oneself was seen as negative. Sometimes students described doing this because it was the easiest route, but other times, students didn't want to risk their group members suffering the consequences if they got the answer wrong.

#### *b. Positive Experiences*

Students reported feeling positively about their peers if they realized they were not alone in their struggles. For a few students, this realization was their greatest success during the semester. They also reported feeling confidence and satisfaction from being able to help other students solve a problem or understand a concept, often citing this as a marker of their own conceptual understanding or mastery. On the flipside, students on the receiving end of that help (when it was productive) felt that their peers were able to explain things to them in a way that their instructors could not because their peers were better able to see the material and/or problem from the perspective of a student and describe things on their level. When students worked together productively, they talked about how comparing their work helped them detect errors and discover new approaches. Students spoke positively about working towards a common goal, being able to contribute without fear of judgment, and being able to rely on others when they felt less confident (although over-reliance was experienced as negative). Finally, students felt motivated when they engaged in friendly competition with others.

## APPENDIX D: RESULTS FOR CODES NOT INCLUDED IN THE MAIN RESULTS

### 1. Social Comparison – a subset of Peers

There were 85 total segments (51.8% of all PEERS segments) coded as both PEERS and SOCIAL COMPARISON. Here, we describe the negative and positive experiences with peers for this SOCIAL COMPARISON subset. Figure D1 below shows differences in the affective valence distributions by whether or not social comparison was reported. 70.4% of the reported negative experiences with peers involved some sort of social comparison, compared to 37.5% of the positive experiences. While this pattern suggests that social comparison is more likely to be salient when reporting negative peer experiences, social comparison was still present in a substantial portion of the positive experiences.

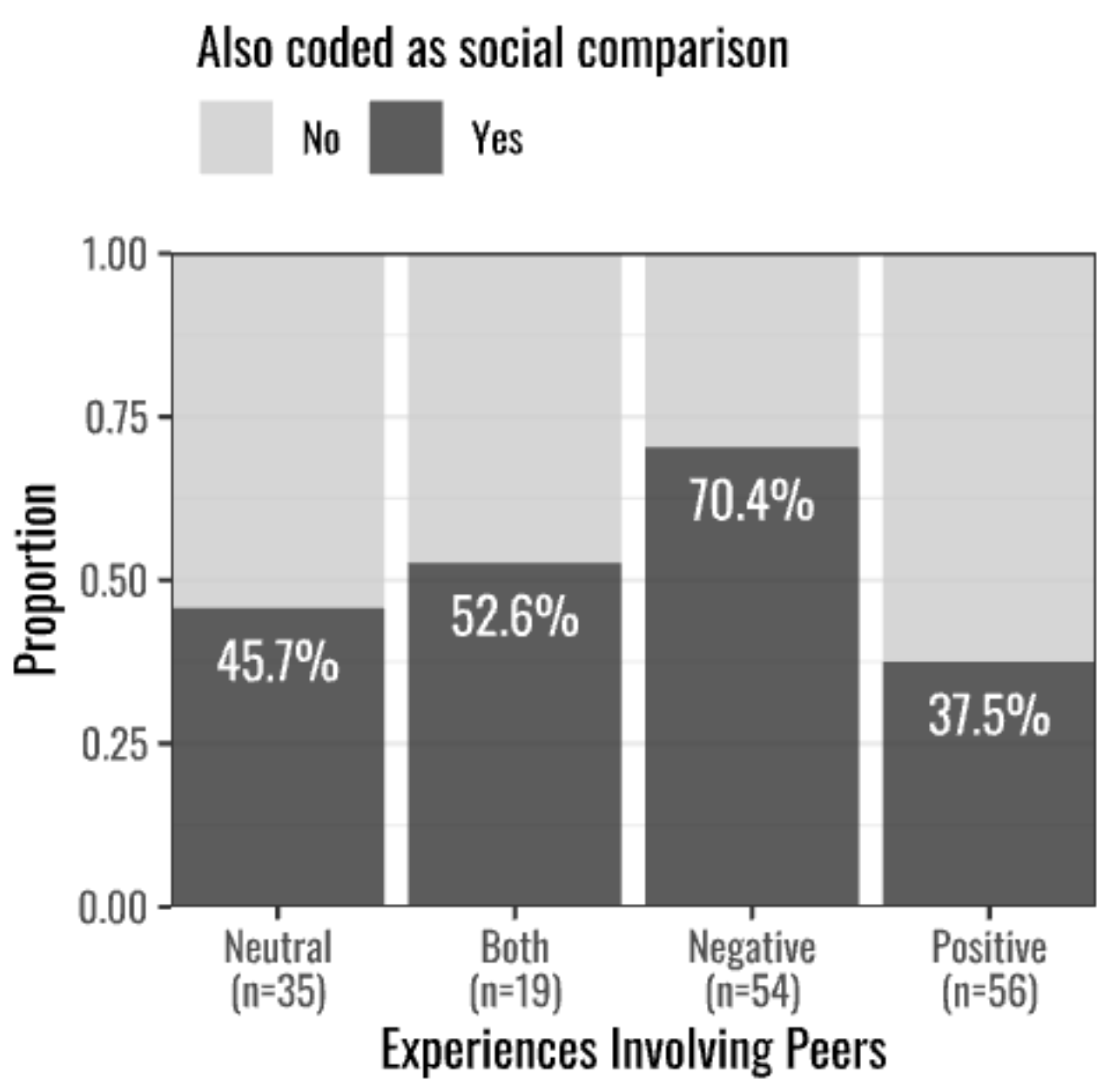

FIG. D1. Experiences with peers involving social comparison.

### 2. Success

There were 143 total segments coded as SUCCESS. Thirty-seven of those were coded only as SUCCESS and so were not analyzed for positive and negative themes. These were primarily responses to question nine from the interview protocol (see TABLE III), when students were asked to define their conception of success in physics. Many students defined success in terms of reaching a standard of performance – achieving a target grade or passing the class – or in terms of mastery – achieving a target level of knowledge and competence with respect to the material [80]. However, these definitions were not mutually exclusive, and some students described them as being context dependent. For example, one student who had taken both courses in the introductory physics series, reported defining success by grades during the first course because they thought they wouldn't need to use the material in the future, and then defining success by mastery in the second course because the material could be directly applied to their other interests. Finally, some students defined success outside of the performance vs. mastery framework all together. Some examples of these definitions are the ability to be comfortable with oneself and one's effort in the face of performance setbacks, the ability to open up to others and reach out for help, and the ability to apply problem solving skills to any difficult or complicated situation in life.

### 3. Suggestions

There were 55 total segments coded as SUGGESTION. Fifty two of those were also coded for other experience codes that were analyzed for affective valence. Three were coded only as SUGGESTION and so were not analyzed for positive and negative themes. The first two segments came from instances where the students were prompted via follow-up questions to give suggestions to a hypothetical incoming student for how to improve their experience in introductory physics. Both segments were about not feeling too bad about getting poor grades. The last segment was about how their experience would have been better if they didn't have to skip class to go to work.